\documentclass[11pt]{article}
\usepackage{float}
\usepackage{graphicx}
\usepackage{amsmath, amssymb}
\usepackage{booktabs}
\usepackage{authblk}
\usepackage{comment}
\usepackage{subfig}
\usepackage{tabularray}
\usepackage{adjustbox}
\usepackage[margin=1in]{geometry}
\usepackage{lineno, blindtext}
\usepackage{hyperref}
\usepackage[nodisplayskipstretch]{setspace}
\usepackage[sorting=none,maxnames=99,minnames=99]{biblatex}
\renewcommand{\bibfont}{\normalsize}

\usepackage[font=normalsize,labelfont=normalsize]{caption}
\usepackage{subfig}

\newcommand{\FullConference}[1]{%
  \begin{center}
    #1
  \end{center}
}

\title{\bfseries New Physics Searches at the LHC through Event-based Anomaly Detection and Development of ADFilter Web-tool}

\author[1]{\textbf{Wasikul Islam}\thanks{Speaker at Lepton Photon 2025 conference.}}
\affil[1]{Department of Physics, University of Wisconsin, Madison, WI 53706 USA}
\author[2]{\textbf{Sergei Chekanov}}
\author[2]{\textbf{Nicholas Luongo}}

\affil[2]{HEP Division, Argonne National Laboratory, USA}

\date{}

\begin{document}

\maketitle
\vspace{-1.5cm} 
\FullConference{%
\centering{
  (Presented at the 32nd International Symposium on Lepton Photon Interactions at High Energies, 25-29 August 2025 at Madison, Wisconsin.)\\
  }
}

\abstract{This work presents advancements in model-agnostic searches for new physics at the Large Hadron Collider (LHC) through the application of event-based anomaly detection techniques utilizing unsupervised machine learning. We discuss the advantages of anomaly detection approach, as demonstrated in a recent ATLAS analysis, and introduce ADFilter, a web-based tool designed to process collision events using autoencoders based on deep unsupervised neural networks. ADFilter calculates loss distributions for input events, aiding in determining the degree to which events can be considered anomalous. Real-life examples are provided to demonstrate how the tool can be used to reinterpret existing LHC results, with the goal of significantly improving exclusion limits. Furthermore, we present a comparative study between anomaly detection and supervised machine learning techniques, using the search for heavy resonances decaying into two or more Higgs bosons as a representative case to demonstrate the application and effectiveness of these methods.}

\vspace{0.1cm} 
{\small Preprint: HEP-ANL-201033}
\vspace{-0.3cm} 

%% \tableofcontents

%\linenumbers

%\maketitle
\thispagestyle{plain}
%\thispagestyle{empty}
%\maketitle
%\clearpage

\pagenumbering{arabic}\setcounter{page}{1}  
\section{Introduction}

The search for physics beyond the Standard Model (BSM) at the LHC has traditionally relied on targeted, model-driven analyses that optimize sensitivity for specific decay topologies. While powerful, such searches may lose coverage for unconventional, unexpected, or poorly modeled signatures. This limitation has motivated the development of \emph{model-agnostic} strategies whose aim is to identify unusual events directly from data without relying on explicit signal hypotheses.

Event-level anomaly detection (AD), implemented using unsupervised autoencoders trained on data, has emerged as a promising framework in this direction. Unlike supervised approaches that require labeled signal samples, anomaly detection relies solely on Standard-Model–like events for training, thus providing a bias-minimized method for uncovering anomalous topologies. These techniques learn the high-dimensional correlations among reconstructed observables and assign an “anomaly score” derived from the reconstruction loss or latent-space structure.

A recent ATLAS study demonstrated the first full Run~2 anomaly-detection search based on deep autoencoders trained using 1287 observables constructed from jets, leptons, photons, and missing transverse momentum~\cite{ATLAS:2023ixc}. The outcome revealed both strong background modeling and enhanced sensitivity to a wide variety of possible new-physics topologies. Motivated by these developments, a publicly accessible web-based reinterpretation framework, \textsc{ADFilter}, has been developed,  which applies the ATLAS autoencoder configuration to user-defined simulated samples. This work summarizes the anomaly-detection methodology, describes the \textsc{ADFilter} framework, and presents comparative studies between anomaly detection and supervised classifiers in the context of di-Higgs resonance searches.

\section{Event-Level Anomaly Detection in ATLAS}

The ATLAS anomaly-detection analysis~\cite{ATLAS:2023ixc} employs a deep autoencoder trained on approximately 1\% of Run~2 proton–proton collision data. Events are represented using the \emph{Rapidity–Mass Matrix} (RMM), a 36\,$\times$\,36 structured array (containing 1287 entries, excluding 9 invariant mass terms) encoding pairwise rapidity differences, invariant masses, and transverse momenta of reconstructed objects. This representation provides a dense summary of the full event topology while maintaining a uniform architecture for all final states.

The autoencoder compresses the 1287-dimensional RMM input into a 200-dimensional latent space before reconstructing it back to the original dimension. Events with large reconstruction loss are classified as anomalous. Thresholds in loss space corresponding to effective signal cross sections of roughly 10~pb, 1~pb, and 0.1~pb define three anomaly regions that bracket increasingly rare topologies.

Using these regions, the analysis performs bump-hunt searches in nine invariant-mass variables: $m_{jj}$, $m_{jb}$, $m_{bb}$, $m_{j\ell}$, $m_{j\gamma}$, $m_{be}$, $m_{b\mu}$,$m_{j\mu}$,$m_{j\gamma}$.
Across all final states considered, no statistically significant deviations from the Standard Model expectation were observed. Localized fluctuations (such as $2.9\sigma$ near 4.8~TeV and $2.8\sigma$ near 1.2~TeV) were found to be consistent with the look-elsewhere effect. Nevertheless, the use of anomaly regions substantially improved sensitivity in the sub-TeV regime, with limits strengthened by factors of $2$–$3$ relative to earlier inclusive strategies. These findings highlight the capability of event-level anomaly detection to complement targeted searches.

\section{The \textsc{ADFilter} Web Tool}

The \textsc{ADFilter} framework~\cite{ADFilter_webpage, Chekanov:2024ezm} was designed to provide the high-energy physics community with a practical mechanism for evaluating anomaly scores for arbitrary simulated models. The tool reproduces the ATLAS anomaly-detection pipeline by implementing the same RMM-based feature construction and the same publicly available trained autoencoder model used in Ref.~\cite{ATLAS:2023ixc}.

\begin{figure}[h] 
\begin{center} \includegraphics[height=0.32\textheight, width=0.85\textwidth]{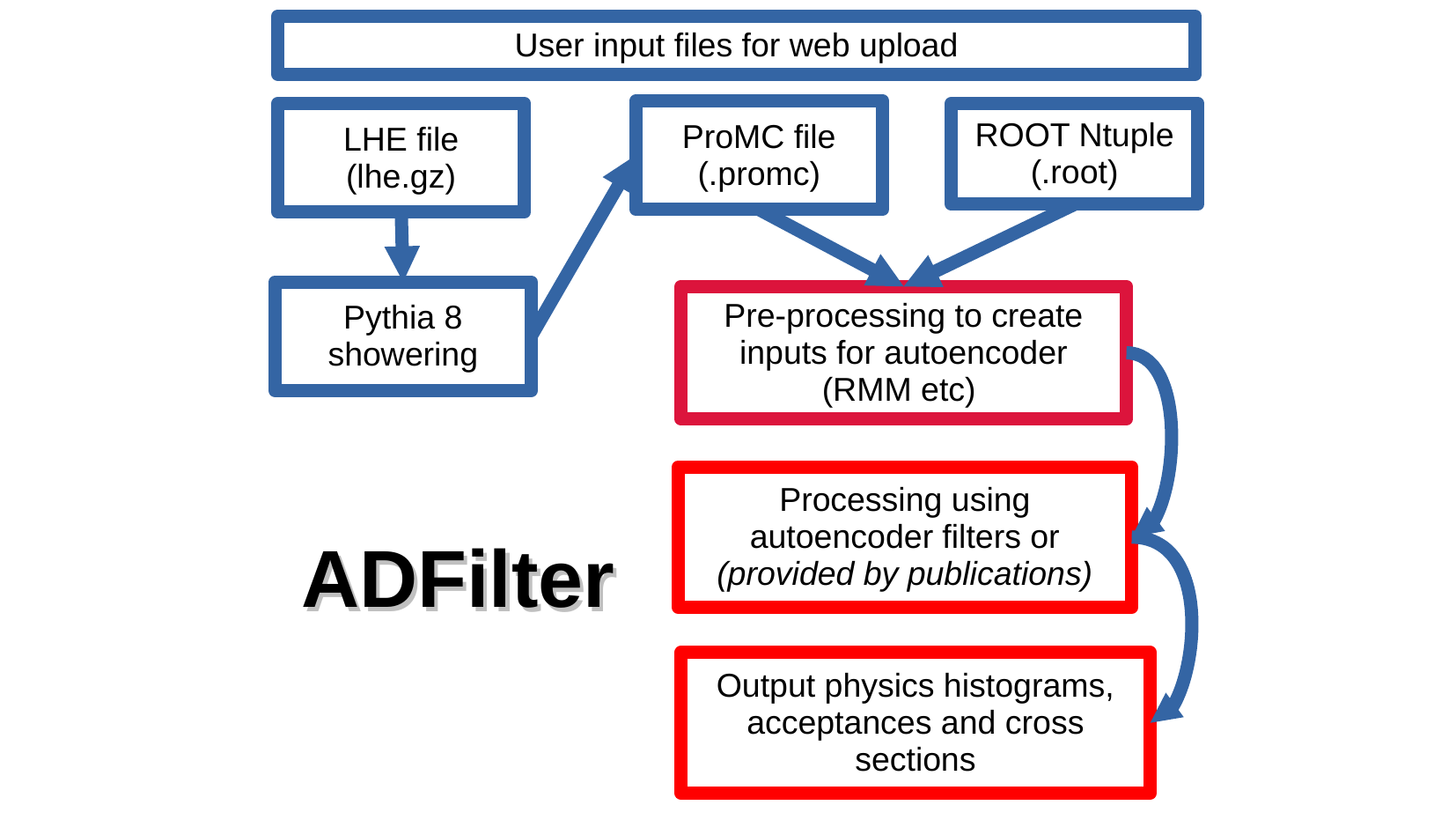} 
\end{center} 
\caption{A schematic diagram showing the workflow of the ADFilter.}  
\label{userGUIwork} 
\end{figure}

\subsection{Workflow}

The \textsc{ADFilter} workflow proceeds as follows:
\begin{enumerate}
    \item Users upload simulated events generated in LHE, HEPMC, ProMC, or Delphes ROOT formats at the ADFilter web interface~\cite{ADFilter_webpage}.
    \item Events are preprocessed and reconstructed into the 1287-dimensional RMM feature set.
    \item The pre-trained autoencoder evaluates each event to compute a reconstruction-loss value.
    \item The anomaly-score distribution is produced, and the fraction of events falling into the ATLAS anomaly regions is reported.
    \item From these results, the \emph{anomaly-detection acceptance} $A$ is computed as
    \[
    A = \frac{N_{\mathrm{AR}}}{N_{\mathrm{tot}}},
    \]
    where $N_{\mathrm{AR}}$ is the number of events inside a chosen anomaly region and $N_{\mathrm{tot}}$ is the total number of generated events.
\end{enumerate}

The acceptance $A$ plays a critical role in reinterpretation of ATLAS limits.  
The model-independent ATLAS anomaly-detection search sets limits on $\sigma \times B \times A \times \epsilon$,
where $\epsilon$ denotes detector and selection efficiency.  
For theorists wishing to compare their BSM models with these limits, knowledge of $A$ is essential. However, $A$ cannot be determined without running the trained autoencoder used in Ref.~\cite{ATLAS:2023ixc}. This requires access to the network weights, RMM construction, and ML infrastructure—elements that are nontrivial for many groups to reproduce.

\textsc{ADFilter} removes this barrier by providing a direct evaluation of $A$ for arbitrary signal hypotheses. A small acceptance ($A \ll 1$) indicates that the BSM model is too similar to the SM background to populate the anomaly regions, while a large acceptance ($A \sim 1$) implies that the model strongly populates anomalous parts of phase space. Once $A$ is known, users can rescale the limits to obtain updated constraints on $\sigma \times B$ for their specific model. This provides a fast and robust route for model reinterpretation using the ATLAS anomaly-detection results.

\subsection{Examples and Use Cases}

Investigations using \textsc{ADFilter} demonstrate its usefulness across a range of benchmark new-physics models~\cite{Chekanov:2024ezm}. Examples include:
\begin{itemize}
    \item A 2~TeV radion scenario, where an enhanced density of high-loss events reflects the more complex jet structure of radion decays.
    \item Heavy vector-boson models ($W'$ and $Z'$), where elevated anomaly scores correspond to the expected kinematic tails of high-mass resonances.
    \item Charged-Higgs production ($tbH^{+}$), where boosted topologies lead to characteristic shapes in the anomaly-loss spectrum and modified exclusion limits.
\end{itemize}

\begin{figure}[h]
  \begin{center}
    \subfloat[Published LHC limits \cite{ATLAS2020}.]{\includegraphics[height=0.29\textheight,width=0.49\textwidth]{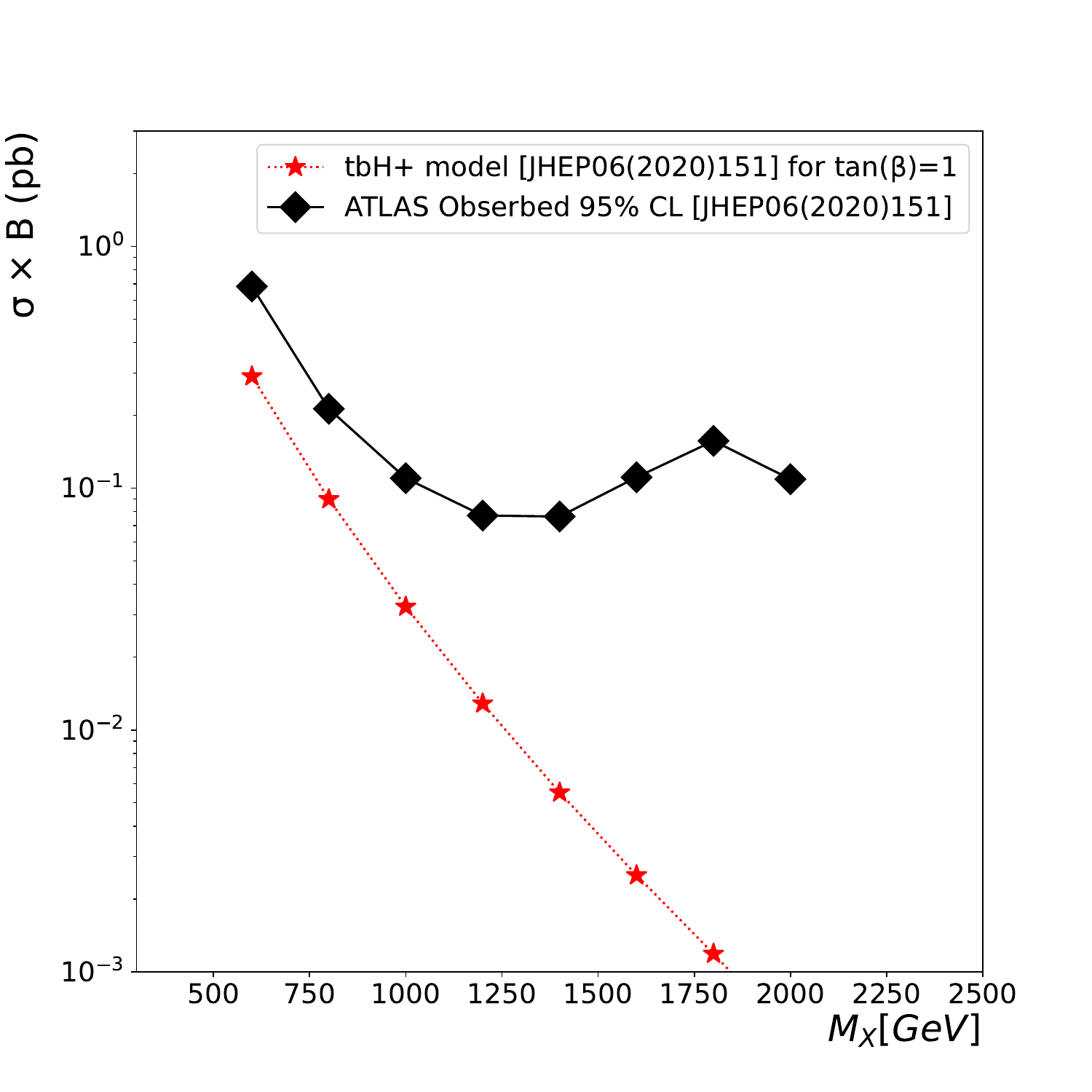}}
    \subfloat[Reinterpreted limits using Ref.\,\cite{ATLAS:2023ixc} and ADFilter.]{\includegraphics[height=0.29\textheight,width=0.49\textwidth]{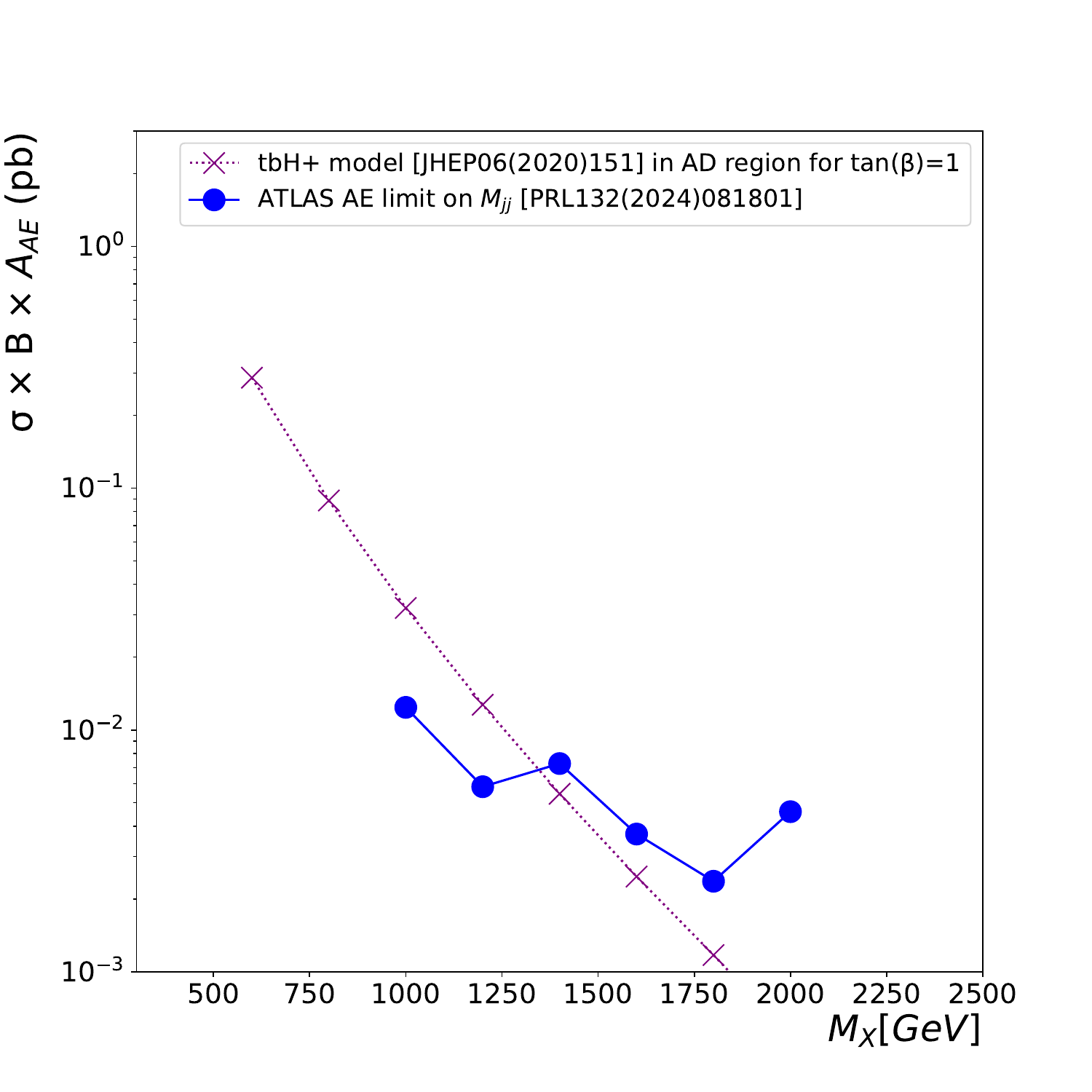}}
  \end{center}
\caption{(a) Observed 95\% credibility-level limits \cite{ATLAS2020} on $\sigma \times B$ for $tbH^{+}$ production as a function of $M_X$, using dijet masses in events with at least one isolated lepton with $p_{T}>60$~GeV.  
(b) ATLAS 95\% limits for Gaussian signals from Ref.~\cite{ATLAS:2023ixc}, rescaled using $(1/A_{\rm sel}\times\varepsilon)$ following Ref.~\cite{ATLAS2020}. 
Magenta points show the cross section after the ADFilter correction. The anomaly-detection interpretation excludes the $\tan\beta=1$ benchmark up to 1.35~TeV, whereas the original selection could not.}
\label{fig:Hplus_2TeV}
\end{figure}

Figure~\ref{fig:Hplus_2TeV} shows, ADFilter showcases significant improvements in limits for for $tbH^{+}$ model using the anomaly detection technique. These studies demonstrate that ADFilter can strengthen exclusion limits and uncover parameter regions that are challenging for more traditional approaches.

\section{Supervised Learning vs.\ Anomaly Detection for di-Higgs}

Further investigations~\cite{Chekanov:2025xpk} explore the comparative behavior of supervised machine-learning classifiers and anomaly-detection models in di-Higgs resonance searches. The scenarios considered include $X\to HH$ and mixed $X\to SH$ decays (with $S\to HH$), which generate rich topologies containing multiple $b$-jets, boosted Higgs bosons, and high-$p_T$ jets.

\begin{figure}[h]
  \begin{center}
    \subfloat[$X \to HH$]{\includegraphics[height=0.28\textheight,width=0.49\textwidth]{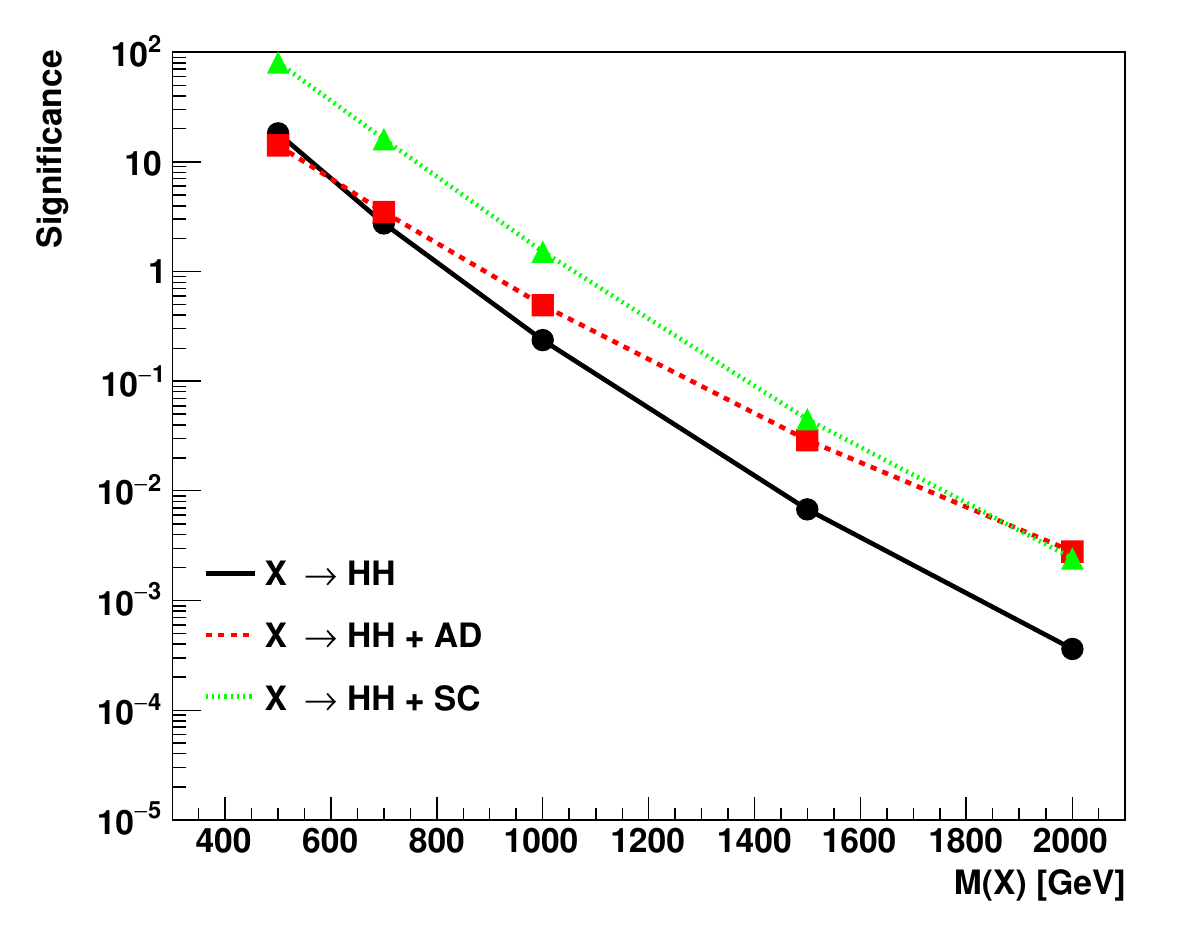}} 
    \subfloat[$X \to SH$]{\includegraphics[height=0.28\textheight,width=0.49\textwidth]{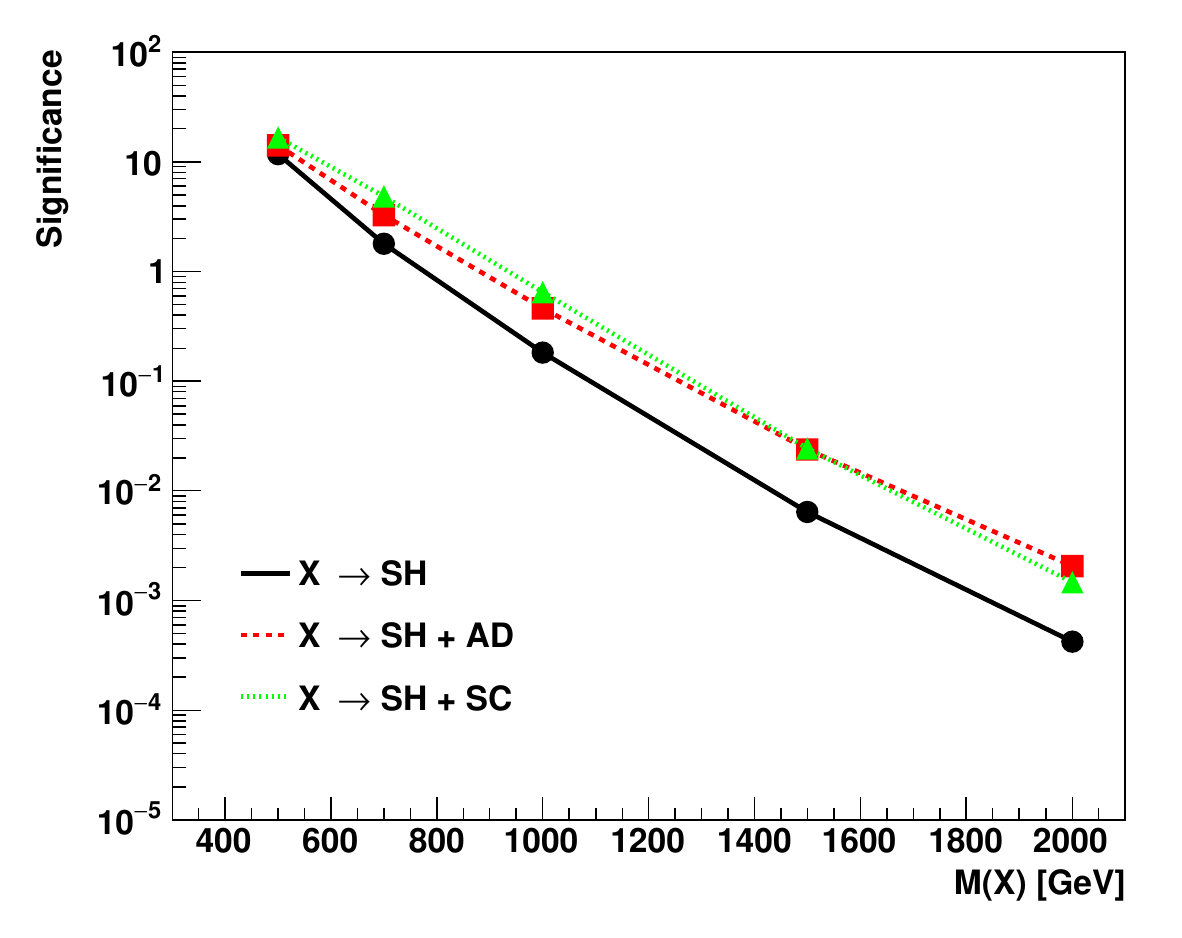}}
  \end{center}
\caption{
Comparison of significance values for 
(a) $X\rightarrow HH$ and 
(b) $X\rightarrow SH$ with $S\rightarrow HH$ for both supervised cluster method (SC) and unsupervised machine learning based anomaly detection(AD) method.
The definitions of signal and background yields, as well as the treatment of uncertainties, follow Ref.~\cite{Chekanov:2025xpk}.
}
\label{fig:signi_cxSM}
\end{figure}

As Fig~\ref{fig:signi_cxSM} shows, the supervised classifiers, trained on labeled signal and background samples, generally achieve the strongest separation when the assumed signal closely matches the training hypothesis. These networks perform particularly well for lower-mass resonances or for decay patterns that resemble documented signatures.

In contrast, anomaly-detection methods do not rely on explicit signal hypotheses and instead focus on deviations from the Standard Model event distribution. This confers an advantage in cases where the signal differs substantially from expected topologies or when the resonance mass becomes large enough that its decay products produce highly non-standard kinematic configurations.

Studies also show that the RMM representation captures characteristic features of di-Higgs events, such as dense substructure patterns from boosted Higgs bosons. These properties naturally influence the anomaly scores, while supervised classifiers benefit from the systematic presence of recurring features in the training data. 

%Studies at Latent-space indicate that anomaly-detection models provide complementary coverage relative to supervised approaches.

\section{Outlook}

Event-level anomaly detection has emerged as a versatile and model-agnostic strategy for exploring new-physics signals at the LHC. The ATLAS search demonstrates both competitive sensitivity and enhanced reach for unexpected signatures. The \textsc{ADFilter} tool makes such techniques broadly accessible, enabling rapid recasting studies for a variety of BSM scenarios.

Future improvements in ADFilter web tool may include the incorporation of additional neural network architectures, support for various detector formats, and the collection of trained anomaly detection models from different detectors and experiments. Combining supervised classifiers, anomaly-detection frameworks, and generative models may also provide a robust pipeline for next-generation searches for new physics across different particle physics experiments.

 \section*{Acknowledgments}
We thank the organizers of the Lepton–Photon 2025 conference. WI is supported by the U.S. Department of Energy (DOE) grant DE-SC0017647. SC and NL are supported by UChicago Argonne, LLC, operator of Argonne National Laboratory, under Contract No. DE-AC02-06CH11357 with the U.S. DOE. Argonne’s work is supported by the DOE Office of High Energy Physics under the same contract. We also acknowledge the computing resources provided by the Laboratory Computing Resource Center at Argonne National Laboratory.

\renewcommand{\bibfont}{\fontsize{11pt}{11pt}\selectfont}
\printbibliography

@article{Chekanov:2024ezm,
    author = "Chekanov, Sergei V. and Islam, Wasikul and Zhang, Rui and Luongo, Nicholas",
    title = "{ADFilter{\textemdash}A Web Tool for New Physics Searches with Autoencoder-Based Anomaly Detection Using Deep Unsupervised Neural Networks}",
    eprint = "2409.03065",
    archivePrefix = "arXiv",
    primaryClass = "hep-ph",
    reportNumber = "ANL-HEP-190964",
    doi = "10.3390/info16040258",
    journal = "Information",
    volume = "16",
    number = "4",
    pages = "258",
    year = "2025"
}

@article{Chekanov:2025xpk,
    author = "Chekanov, Sergei V. and Islam, Wasikul and Luongo, Nicholas",
    title = "{Enhancing Sensitivity for Di-Higgs Boson Searches Using Anomaly Detection and Supervised Machine Learning Techniques}",
    eprint = "2504.12418",
    archivePrefix = "arXiv",
    primaryClass = "hep-ph",
    reportNumber = "HEP-ANL-195813",
    month = "4",
    year = "2025"
}

@other{ADFilter_webpage,
    title = "{ADFilter webpage}",
    url = "https://mc.hep.anl.gov/adfilter"
}

@article{ATLAS2020,
	author = "ATLAS Collaboration",
	collaboration = "ATLAS",
	title = "{Search for dijet resonances in events with an isolated charged lepton using $\sqrt{s} = 13$ TeV proton-proton collision data collected by the ATLAS detector}",
	eprint = "2002.11325",
	archivePrefix = "arXiv",
	primaryClass = "hep-ex",
	reportNumber = "CERN-EP-2019-276",
	doi = "10.1007/JHEP06(2020)151",
	journal = "JHEP",
	volume = "06",
	pages = "151",
	year = "2020"
}

@article{ATLAS:2023ixc,
    author = "ATLAS Collaboration",
    collaboration = "ATLAS",
    title = "{Search for New Phenomena in Two-Body Invariant Mass Distributions Using Unsupervised Machine Learning for Anomaly Detection at $\sqrt{s}=13$ TeV with the ATLAS Detector}",
    reportNumber = "CERN-EP-2023-112",
    doi = "10.1103/PhysRevLett.132.081801",
    journal = "Phys. Rev. Lett.",
    volume = "132",
    number = "8",
    pages = "081801",
    year = "2024"
}

%\begin{thebibliography}{99}
%\bibitem{Evans:1129806}
%\bibitem{Aad:2008zzm}
%\bibitem{Aad:2020kep}
%\bibitem{ATLAS-CONF-2011-069}
%\bibitem{Aaboud:2018cwk}
%....
%
%\end{thebibliography}

%\usepackage[square,numbers]{natbib}
%\bibliographystyle{abbrvnat}

%\bibliographystyle{unsrt}

\end{document}